\pdfoutput=1
\documentclass[conference]{IEEEtran}
\usepackage{graphicx} % Required for inserting images
\usepackage[percent]{overpic}
\usepackage{cite}
\usepackage[colorlinks=true, linkcolor=black, citecolor=black, urlcolor=blue, bookmarks=false]{hyperref}
\usepackage{amsmath, amsfonts, amssymb}
\usepackage{algorithm}
\usepackage{algpseudocode}
\usepackage{xcolor}
\usepackage{colortbl}
\newcommand{\revb}[1]{#1}
\usepackage{caption}
\newtheorem{proposition}{Proposition}
\newtheorem{remark}{Remark}

\IEEEoverridecommandlockouts
\IEEEaftertitletext{\vspace{-1\baselineskip}}

\title{Self-Supervised Operator Learning for\\
Channel Estimation in Fluid Antenna Systems\vspace{-12pt}}

\author{
    \IEEEauthorblockN{
        Erqiang Tang,
        Zixin Wang,
        Wei Guo,
        Shenghui Song,
        Jun Zhang, \textit{Fellow, IEEE},
        Khaled B. Letaief, \textit{Fellow, IEEE}
    }
    \IEEEauthorblockA{
        \revb{Dept. of ECE}, The Hong Kong University of Science and Technology, Hong Kong \\
        Emails: etangaa@connect.ust.hk, \revb{\{eewangzx, eeweiguo, eeshsong, eejzhang, eekhaled\}@ust.hk}
    }
}

\begin{document}

\maketitle

\begin{abstract}
\revb{Fluid antenna systems (FAS) have emerged as a promising technology for enhancing wireless communications through adaptive antenna repositioning. Recently, Fourier neural operators (FNOs) have shown strong potential for FAS channel estimation owing to their ability to learn resolution-invariant mappings and generalize across different port densities. However, existing operator learning approaches require densely sampled channel realizations as training labels, whereas a practical receiver only collects noisy pilot observations from a small subset of ports. This mismatch between training requirements and practical channel acquisition poses a major obstacle to deployment. To address this challenge, we develop a self-supervised framework that trains the FNO-based estimator directly from noisy sub-sampled pilots without any labeled channel fields. The proposed objective combines a pilot splitting strategy, which creates supervision at ports hidden from the model input, with a correction based on Stein's unbiased risk estimate (SURE) that eliminates the measurement noise bias. We prove that this objective admits the same minimum mean square error (MMSE) solution as the ideal supervised criterion.} Since the loss relies solely on received pilots, the estimator can be continually adapted online \revb{on the live pilot stream}. Simulation results show that the self-supervised FNO \revb{achieves accuracy comparable to supervised training}, generalizes zero-shot to unseen \revb{port resolutions}, and rapidly recovers from \revb{distribution shift} via online adaptation.
\end{abstract}


\section{Introduction}\label{sec:intro}

The fluid antenna system (FAS) has emerged as a transformative technology to circumvent the fundamental limits of fixed-geometry arrays in exploiting spatial degrees of freedom (DoF)~\cite{wong2020fluid,zhu2024historical}. By allowing antennas to adaptively reposition themselves within a prescribed space, FAS significantly improves the utilization of spatial DoF without escalating hardware complexity. \revb{Nonetheless}, optimal antenna positioning, \revb{i.e., port selection,} requires complete channel state information (CSI) across all candidate ports, whereas only a fraction of them can be observed within one coherence block, making FAS channel estimation a severely ill-posed inverse problem.

Leveraging the inherent spatial sparsity of the propagation environment, early efforts in FAS channel estimation typically adopted compressed sensing (CS)-based techniques to recover full CSI from partial port observations~\cite{xiao2024channel,xu2024sparse}. However, this sparsity assumption fails in rich-scattering environments, leading to significant performance degradation. Furthermore, the iterative nature of these algorithms incurs considerable computational complexity that scales poorly with increasing port resolution. These limitations have motivated learning-based estimators, which capture complex spatial channel structures directly from data~\cite{tang2025accurate} and bypass iterative computation via fast feedforward channel extrapolation~\cite{zhang2024agmae}. Despite these advantages, such estimators are built on a fixed discrete port grid, which ties the model to the specific port resolution used during training. Consequently, generalizing to a different port density at deployment will induce a \revb{\emph{dimension mismatch}} problem, under which the estimation performance is no longer guaranteed. Mitigating this typically requires costly retraining for each new grid. \revb{A recent study employs self-supervised masked reconstruction to reduce the labeling burden~\cite{gao2025ssnet}, but its training still relies on clean channel samples from simulators and remains tied to a fixed port grid.}

Recently, the pioneering work~\cite{wang2026rethinking} addressed the \revb{dimension mismatch} issue by recasting FAS channel estimation as an operator learning task. Specifically, it employs a Fourier neural operator (FNO)~\cite{li2021fourier} to learn a resolution-invariant, field-level mapping through a parameterization that is independent of the discrete spatial grid, so that a single trained model can generalize zero-shot across various port densities. However, this neural operator-based estimator is trained via offline supervised regression \revb{and thus still requires} clean full-field channel samples as training labels. In practice, such labels are inaccessible as the receiver only observes noisy pilots over a sparse subset of ports. \revb{This creates a fundamental deployment gap, where the training requirements of operator learning are incompatible with practical channel acquisition, so} the model must \revb{instead} be trained on synthetic data generated by an offline simulator. Therefore, whenever the simulated channel statistics deviate from the true propagation environment, the learned estimator suffers from a severe \revb{\emph{distribution shift}}, \revb{which cannot be corrected online since the live pilot stream provides no clean labels}. \revb{To the best of our knowledge, this challenge has not been addressed in the FAS literature, as it requires handling port sub-sampling and measurement noise jointly.}

In this paper, we \revb{establish the first self-supervised operator learning framework for FAS channel estimation that} learns the estimator solely from the noisy sub-sampled pilots available at the receiver\revb{, thereby bridging the deployment gap between operator learning and practical channel acquisition}. \revb{To create supervision at ports unseen by the model, we introduce a \emph{pilot splitting} strategy inspired by Ambient Diffusion~\cite{daras2023ambient}, which partitions the observed ports into a visible set serving as the model input and a masked set scored by the loss, so that randomizing the split across blocks extends supervision to the entire port grid.} \revb{Furthermore, we incorporate a correction based on Stein's unbiased risk estimate (SURE)~\cite{stein1981estimation} to compensate for measurement noise, and prove that the resulting loss is minimized by the minimum mean square error (MMSE) estimator induced by the true channel distribution, as is the ideal supervised objective, establishing a principled foundation for self-supervised operator learning in FAS.} \revb{Since the loss depends on the received pilots alone, the estimator can be continually updated on the live pilot stream, so that it directly learns the channel distribution of the deployment environment and thereby mitigates the distribution shift.} Simulation results demonstrate that \revb{the self-supervised FNO achieves estimation accuracy on par with its supervised counterpart}, while maintaining zero-shot generalization to unseen port densities.

\section{System Model and Problem Formulation}\label{sec:model}

In this section, we first present the continuous channel model for the considered two-dimensional (2D) FAS. Then, we introduce the pilot signal model and formulate the channel estimation problem as a field-level estimation task to achieve resolution-invariant estimation.

\subsection{Channel Model}\label{sec:channel}

\begin{figure}[t]
    \centering
    \includegraphics[width=0.81\columnwidth,trim=0 16 0 16,clip]{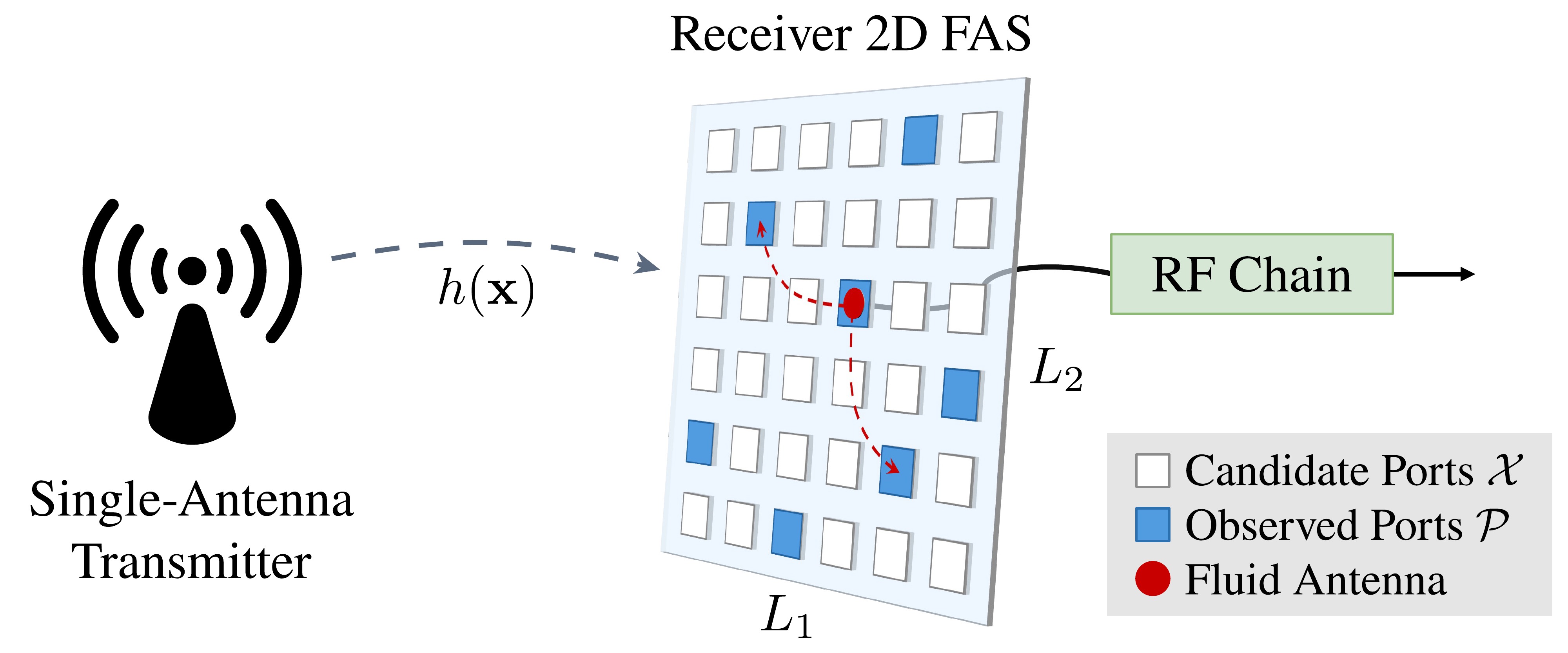}
    \caption{\revb{Illustration of the considered 2D FAS system.}}
    \label{fig:system}
\end{figure}

\revb{As shown in Fig.~\ref{fig:system}, we} consider a point-to-point communication system comprising a transmitter with a single fixed-position antenna and a receiver equipped with a 2D FAS\revb{, which consists of a single fluid antenna element connected to one radio frequency (RF) chain}. The FAS surface occupies a rectangular aperture $\mathcal{D}=[0,L_{1}]\times[0,L_{2}]\subset\mathbb{R}^{2}$, which is uniformly discretized into an $N_{1}\times N_{2}$ grid of $N = N_1 N_2$ candidate ports $\mathcal{X}=\{\mathbf{x}_{1},\dots,\mathbf{x}_{N}\}\subset\mathcal{D}$, with $\mathbf{x}_{n}=[x_{n,1},x_{n,2}]^{\top}$ denoting the spatial coordinates of the $n$-th port. At any time instant, the fluid antenna can be switched to any of these ports. The port density may vary across different deployments.

Let $h(\mathbf{x})\in\mathbb{C}$ denote the complex channel coefficient at an arbitrary position $\mathbf{x} \in \mathcal{D}$. To better capture diffuse scattering in practical propagation environments, we adopt a continuous angular scattering channel model. Under the narrowband far-field assumption, the electromagnetic field across the FAS aperture $\mathcal{D}$ can be decomposed into a superposition of plane waves, characterized by their directions of arrival. Specifically, let $\theta, \phi \in[-\pi/2,\pi/2]$ represent the elevation and azimuth angles of arrival (AoAs), respectively. The channel $h(\mathbf{x})$ is then given by\footnote{This formulation subsumes the discrete multipath model as a special case by setting $\rho(\mathbf{u}) = \sum_{i=1}^{N_p} g_i \delta(\mathbf{u} - \mathbf{u}_i)$, where $g_i$ and $\mathbf{u}_i$ are the complex gain and virtual AoA of the $i$-th path.}
\begin{equation}\label{eq:cont-channel}
    h(\mathbf{x}) = \int_{-\sin\Theta}^{\sin\Theta} \int_{-\sin\Phi\sqrt{1-u_{2}^{2}}}^{\sin\Phi\sqrt{1-u_{2}^{2}}}\rho(\mathbf{u})\,e^{-j\frac{2\pi}{\lambda}\mathbf{u}^{\top}\mathbf{x}}\,du_{1}\,du_{2},
\end{equation}
where $\lambda$ is the carrier wavelength, $\mathbf{u}=[u_{1},u_{2}]^{\top}$ is the virtual AoA vector with $u_{1}=\cos\theta\sin\phi$ and $u_{2}=\sin\theta$, $\rho(\mathbf{u})$ denotes the \emph{angular scattering density function}, and $\Theta, \Phi$ are the maximum half angular spreads in elevation and azimuth, respectively, \revb{so that $|\theta|\leq\Theta$ and $|\phi|\leq\Phi$}.

\subsection{Signal Model and Problem Formulation}\label{sec:problem}

We assume a block-fading channel that remains static within each coherence block, during which the affordable pilot overhead allows the receiver to collect only $M \ll N$ port observations, forming a sampling pattern $\mathcal{P}\subset\mathcal{X}$ with $|\mathcal{P}| = M$\revb{, as highlighted in Fig.~\ref{fig:system}}. Switching the fluid antenna across $\mathcal{P}$ yields the observations
\begin{equation}\label{eq:obs}
    y_{n}=h(\mathbf{x}_{n})+w_{n},\qquad \mathbf{x}_{n}\in\mathcal{P},
\end{equation}
where $w_{n} \sim \mathcal{CN}(0,\sigma_w^2)$ represents the independent and identically distributed (i.i.d.) additive white Gaussian noise (AWGN), whose variance $\sigma_w^2$ is known at the receiver. The sampling pattern $\mathcal{P}$ is selected by the receiver independently of the channel realization and varies across blocks. For notational brevity, the resulting $M$ observations in~\eqref{eq:obs} are collectively denoted by $\{y_n\}$.

To facilitate downstream port selection, the channel estimation task aims to recover $h(\mathbf{x}_n)$ for all $\mathbf{x}_n \in \mathcal{X}$ from the noisy sub-sampled observations $\{y_n\}$, the sampling pattern $\mathcal{P}$, and the noise level $\sigma_w^2$, which is inherently an ill-posed inverse problem. \revb{Traditional learning-based methods formulate this task as a discrete mapping from the $M$ observations to the $N$ port coefficients, which ties the trained model to the specific grid seen during training, so that the estimation performance is no longer guaranteed once the spatial resolution changes.} To eliminate this grid dependency and achieve resolution-invariant estimation, we recast channel estimation as a field-level mapping problem:
\begin{equation}\label{eq:estimator}
    \mathcal{G}:\big(\{y_n\},\mathcal{P},\sigma_{w}^2\big)\mapsto\hat{h}(\cdot).
\end{equation}
\revb{To this end, we propose an operator learning framework based on FNOs to directly learn $\mathcal{G}$ in~\eqref{eq:estimator}. Once the continuous channel field $\hat{h}(\cdot)$ over $\mathcal{D}$ is acquired by the well-trained operator, the channel coefficients of any candidate port set in $\mathcal{D}$ can be directly evaluated via $\{\hat{h}(\mathbf{x}_{n})\}_{n=1}^{N}$, for an arbitrary grid resolution $N_{1}\times N_{2}$. The detailed design of this framework is elaborated in the following section.}

\section{Fourier Neural Operator for Resolution-\\Invariant FAS Channel Estimation}\label{sec:operator}

This section establishes the FNO-based operator learning framework for resolution-invariant FAS channel estimation. First, we detail the adopted model architecture, which learns to recover the underlying continuous channel field $\hat{h}(\cdot)$ independently of spatial grids and enables consistent evaluation across arbitrary port resolutions. We then conduct a spatial spectral analysis of the FAS channel field to demonstrate why the FNO’s mode truncation mechanism is naturally suited for capturing its intrinsic spatial structure.

\subsection{Model Architecture}\label{sec:fno}

To realize the field-level mapping formulated in~\eqref{eq:estimator}, we adopt a U-shaped FNO backbone following the model architecture proposed in~\cite{wang2026rethinking}, which sequentially processes the input through a lifting layer, an $L$-stage Fourier encoder-decoder structure \revb{featuring symmetric skip connections and a global residual link}, and an output layer. \revb{This U-shaped hierarchy extracts spatial features at multiple scales, while the skip connections preserve fine-grained details.} \revb{To begin with, we} encode the sub-sampled observations, sampling mask, spatial coordinates, and noise information into a unified grid-based representation. Specifically, the input feature vector $\mathbf{a}_n \in \mathbb{R}^6$ at the $n$-th candidate port $\mathbf{x}_n \in \mathcal{X}$ is constructed as \begin{equation}\label{eq:input-enc}
    \mathbf{a}_{n}=\big[\Re\{\tilde{y}_{n}\},\,\Im\{\tilde{y}_{n}\},\,m_{n},\,\bar{x}_{n,1},\,\bar{x}_{n,2},\,\sigma_{w}^{2}\big]^{\top},
\end{equation}
where \revb{$\Re\{\cdot\}$ and $\Im\{\cdot\}$ denote the real and imaginary parts, respectively,} $\tilde{y}_n$ represents the zero-filled observation \revb{that equals $y_n$ if $\mathbf{x}_n \in \mathcal{P}$ and $0$ otherwise}, $m_n \in \{0,1\}$ is the binary sampling mask with $m_{n}=1$ if and only if $\mathbf{x}_{n}\in\mathcal{P}$, and $\bar{x}_{n,i} = \tfrac{x_{n,i}}{L_i} \in [0,1]$ denotes the normalized spatial coordinate \revb{for $i=1,2$}.

Although the input feature vectors are constructed discretely, every module of the FNO is defined at the continuous function level over the aperture. \revb{In particular, the lifting layer first applies a pointwise affine transformation $v_{0}(\mathbf{x}_{n})=\mathbf{W}_{\mathrm{up}}\mathbf{a}_{n}+\mathbf{b}_{\mathrm{up}}$ with $\mathbf{W}_{\mathrm{up}}\in\mathbb{R}^{d\times6}$ and $\mathbf{b}_{\mathrm{up}}\in\mathbb{R}^{d}$, which embeds each input vector into a $d$-dimensional latent representation to provide sufficient feature capacity for the subsequent Fourier layers. Since the same transformation is shared across all positions, it conceptually defines a feature field $v_{0}$ over the entire aperture $\mathcal{D}$, and the lifted vectors are simply its discrete samples at the ports $\mathbf{x}_{n}$.} The feature field is then iteratively updated by $2L$ Fourier layers across the encoder and decoder, where the $\ell$-th layer is defined as
\begin{equation}\label{eq:fourier-layer}
    v_{\ell+1}(\mathbf{x})=\eta\Big(\mathbf{W}_{\ell}v_{\ell}(\mathbf{x})+\mathcal{F}^{-1}\big(\mathbf{R}_{\ell}\cdot\mathcal{F}[v_{\ell}]\big)(\mathbf{x})\Big),\quad\forall\mathbf{x}\in\mathcal{D}.
\end{equation}
Here, $\mathcal{F}$ and $\mathcal{F}^{-1}$ denote the 2D Fourier transform and its inverse, respectively, $\mathbf{W}_{\ell}\in\mathbb{R}^{d\times d}$ is a learnable weight matrix, and $\eta(\cdot)$ is a nonlinear activation function. From a signal processing perspective, the spectral branch $\mathcal{F}^{-1}(\mathbf{R}_{\ell}\cdot\mathcal{F}[v_{\ell}])$ acts as a learnable low-pass spatial filter. Specifically, $\mathbf{R}_{\ell}$ performs \emph{mode truncation} by multiplying the Fourier coefficients of $v_{\ell}$ at the low-frequency modes $\mathbf{k}\in\mathcal{K}\triangleq\{\mathbf{k}\in\mathbb{Z}^{2}:|k_{1}|\leq K_{1},|k_{2}|\leq K_{2}\}$ by a weight matrix $\mathbf{R}_{\ell}(\mathbf{k})\in\mathbb{C}^{d\times d}$, while filtering out all the remaining high-frequency modes. To adaptively suppress measurement noise, it further applies soft-thresholding to each retained coefficient $c$ by shrinking it to $c \cdot \max(0,1-\gamma/|c|)$ with a learnable threshold $\gamma$. By doing so, the spectral branch mainly captures the global spatial dependencies, while the parallel pointwise branch $\mathbf{W}_{\ell}$ preserves local high-frequency details and the nonlinear activation $\eta(\cdot)$ dynamically blends both. In practical implementation, each layer is evaluated on the port grid, where $\mathcal{F}$ can be efficiently computed via the fast Fourier transform (FFT).

Finally, a pointwise multilayer perceptron (MLP) maps the final feature field $v_{2L}(\mathbf{x})$ into two scalars that represent the real and imaginary parts of the continuous channel estimate $\hat{h}(\mathbf{x})$. Evaluated over the candidate port grid, this projection directly outputs the desired channel coefficients $\{\hat{h}(\mathbf{x}_n)\}_{n=1}^N$. Crucially, while the model processes discretized data in practice, its parameterization remains strictly independent of the grid resolution. In particular, all learnable modules consist of either spatial pointwise operations shared across all positions or spectral weights assigned to a fixed mode set $\mathcal{K}$, where the grid setup dictates only the FFT size. Moreover, since each spectral weight $\mathbf{R}_{\ell}(\mathbf{k})$ is learned directly in the Fourier space, evaluating the model on a different grid merely samples the same set of Fourier basis functions at new positions~\cite{li2021fourier}. Consequently, a single set of trained weights characterizes a consistent underlying operator across any $N_1 \times N_2$ resolution, including those unseen during training, thereby endowing the resulting FNO, denoted as $\mathcal{G}_{\boldsymbol{\theta}}:(\{y_{n}\},\mathcal{P},\sigma_{w}^{2})\mapsto\hat{h}(\cdot)$, with the resolution invariance property required by~\eqref{eq:estimator}.

\subsection{Spectral Structure of the 2D FAS Channel Field}\label{sec:spectral}

The natural fit between the FNO framework and FAS channel estimation is rooted in the spatial spectral structure of the channel field. Revisiting the channel model in~\eqref{eq:cont-channel} from the perspective of a 2D spatial inverse Fourier transform, the continuous field $h(\mathbf{x})$ is essentially a superposition of plane-wave components $e^{-j\frac{2\pi}{\lambda}\mathbf{u}^{\top}\mathbf{x}}$, each of which acts as a 2D complex exponential characterized by the spatial frequency vector $\mathbf{f}_s = -\mathbf{u}/\lambda$. Given that $\mathbf{u}$ represents the virtual AoA vector of the corresponding far-field incident wave, its $\ell_2$-norm is bounded as $\Vert{}\mathbf{u}\Vert{} \le 1$. Consequently, even under the extreme condition of full angular coverage \revb{with $\Theta=\Phi=\pi/2$}, the spatial spectrum of the channel field remains strictly confined within a compact disk of radius $1/\lambda$, making the ideal 2D FAS channel inherently band-limited in the spectral domain. While any practical finite aperture $\mathcal{D}$ induces spectral leakage via spatial windowing, \revb{most} of the channel energy remains concentrated within this compact disk.

When applying the FNO over the 2D continuous aperture $\mathcal{D}$, the $\mathbf{k}$-th Fourier mode corresponds to the spatial frequency vector $\big[ \frac{k_1}{L_1}, \frac{k_2}{L_2} \big]^{\top}$. The above band limitation thus implies that the channel energy is predominantly concentrated on the low-frequency modes satisfying $\bigl(\frac{k_1\lambda}{L_1}\bigr)^2 + \bigl(\frac{k_2\lambda}{L_2}\bigr)^2 \le 1$, which covers an elliptical region. Accordingly, the maximum frequency span along each axis is restricted to $\vert{}k_i\vert{} \leq L_i/\lambda$. Therefore, the mode truncation budget $K_i$ for the FNO only needs to be set such that $K_i > L_i/\lambda$ with a modest margin. This ensures the spectral branch efficiently captures the dominant low-frequency energy, while the parallel pointwise branch $\mathbf{W}_\ell$ naturally compensates for the residual high-frequency leakage induced by the finite aperture.

\revb{Notably,} the truncation budget depends exclusively on the aperture size in wavelengths and remains strictly independent of both the port resolution and the specific scattering environment. Thus, a single FNO backbone configured at design time generalizes across deployments with different port densities.

\section{Proposed Self-Supervised Training Framework}\label{sec:gtfree}

\revb{Upon establishing the resolution-invariant FNO architecture, we now propose a self-supervised training framework to optimize and adapt $\mathcal{G}_{\boldsymbol{\theta}}$ from noisy sub-sampled pilots alone.} We first formulate the training loss and show how it enables online adaptation without any labels, and then establish its optimality by proving that it \revb{is minimized by the same MMSE estimator as} the ideal supervised objective.

\subsection{Self-Supervised Loss Construction}\label{sec:loss}

\begin{figure}[t]
    \centering
    \includegraphics[width=0.9\columnwidth]{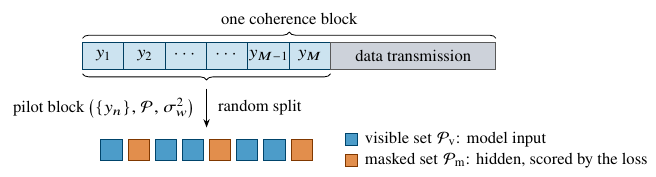}
    \caption{\revb{Illustration of the proposed pilot splitting strategy.}}
    \label{fig:splitting}
\end{figure}

\revb{Training the operator $\mathcal{G}_{\boldsymbol{\theta}}$ conventionally relies on supervised regression over clean full-field channel labels, which are never available in practical FAS deployments, as discussed in Section~\ref{sec:intro}. This motivates a training objective that consumes only the noisy pilots and the noise level known at the receiver.} Specifically, for each coherence block, \revb{the receiver} acquires a set of noisy measurements $\{y_n\}$ governed by the signal model in~\eqref{eq:obs}, subject to \revb{a block-specific noise level $\sigma_w^2$ and sampling pattern $\mathcal{P}$, whose size may also vary across blocks}. \revb{We refer to the resulting per-block triplet $(\{y_n\}, \mathcal{P}, \sigma_w^2)$ as a \emph{pilot block}, as illustrated in Fig.~\ref{fig:splitting}, and the proposed loss will be built solely upon such pilot blocks.}

A straightforward attempt at training without labels is naive self-supervision, where the model takes all available observations as input \revb{while the same noisy observations also serve as the regression targets}, i.e.,
\begin{equation}\label{eq:naive-loss}
    \mathcal{L}_{\mathrm{naive}}(\boldsymbol{\theta})=\mathbb{E}\Big[\textstyle\sum_{\mathbf{x}_{n}\in\mathcal{P}}\big|\hat{h}(\mathbf{x}_{n})-y_{n}\big|^{2}\Big],
\end{equation}
with $\hat{h}(\cdot)=\mathcal{G}_{\boldsymbol{\theta}}(\{y_n\},\mathcal{P},\sigma_{w}^{2})(\cdot)$. This naive loss suffers from two critical flaws. \revb{First, since the ports outside $\mathcal{P}$ are never penalized, the model outputs at those locations can degrade arbitrarily without affecting the loss. Second, the regression targets $y_n$ are corrupted by noise, so that minimizing~\eqref{eq:naive-loss} drives the model to fit the specific noise realizations.}

\revb{The first flaw is addressed by \emph{pilot splitting}, which creates supervision at ports hidden from the model input. Specifically, following the Ambient Diffusion framework~\cite{daras2023ambient}, we randomly partition the sampling pattern $\mathcal{P}$ of each pilot block into a visible set $\mathcal{P}_{\mathrm{v}}$ and a nonempty masked set $\mathcal{P}_{\mathrm{m}}$, where the split is drawn independently of the channel and noise realizations.} During training, the model only sees the visible set $\mathcal{P}_{\mathrm{v}}$, whereas the loss scores the output on the entire $\mathcal{P}$. As a result, the masked ports act as hidden evaluation points, forcing the FNO to infer the channel field beyond its visible inputs. \revb{On top of the split, a diffusion-style noise injection lifts the visible observations} to a common noise level $\sigma_t \ge \sigma_w$ via $z_n = y_n + \sqrt{\sigma_t^2-\sigma_w^2}\,\epsilon_n$ for $\mathbf{x}_n \in \mathcal{P}_{\mathrm{v}}$, with $\epsilon_n \sim \mathcal{CN}(0,1)$ i.i.d. and independent of $w_n$, and $\sigma_{t}$ is drawn from a predefined schedule that includes the unlifted case $\sigma_{t}=\sigma_{w}$. This enables a single trained model to generalize across all noise levels in the schedule, so that pilot blocks with heterogeneous SNRs can share one set of weights.

\revb{The second flaw of noisy regression targets is in turn handled by incorporating a SURE correction into the training objective.} With $\hat{h}_t(\cdot) \triangleq \mathcal{G}_{\boldsymbol{\theta}}\big(\{z_n\},\, \mathcal{P}_{\rm v},\, \sigma_t^2\big)(\cdot)$ denoting the output field, the self-supervised loss is formulated as
\begin{equation}\label{eq:gtfree-loss}
\begin{aligned}
    \mathcal{L}_{\mathrm{prop}}&(\boldsymbol{\theta})=\mathbb{E}\bigg[\sum_{\mathbf{x}_{n}\in\mathcal{P}_{\mathrm{m}}}\big|\hat{h}_{t}(\mathbf{x}_{n})-y_{n}\big|^{2}\\
    &+\sum_{\mathbf{x}_{n}\in\mathcal{P}_{\mathrm{v}}}\Big(\big|\hat{h}_{t}(\mathbf{x}_{n})-y_{n}\big|^{2}+2\sigma_{w}^{2}\,\Re\Big\{\tfrac{\partial\hat{h}_{t}(\mathbf{x}_{n})}{\partial z_{n}}\Big\}\Big)\bigg],
\end{aligned}
\end{equation}
where $\frac{\partial}{\partial z_{n}}\triangleq\frac{1}{2}\big(\frac{\partial}{\partial\Re\{z_{n}\}}-j\frac{\partial}{\partial\Im\{z_{n}\}}\big)$ denotes the complex Wirtinger derivative with respect to the noise-injected input at the corresponding port. Note that the SURE correction term is scaled by the measurement noise variance $\sigma_{w}^{2}$ rather than $\sigma_{t}^{2}$, since only the measurement noise contaminates the regression targets, while the injected noise solely affects the input. In practical implementation, \revb{the sum of derivatives in~\eqref{eq:gtfree-loss}, referred to as the divergence,} is efficiently estimated via a first-order finite difference. Specifically, by perturbing the visible inputs with i.i.d. $b_n$ drawn uniformly from \revb{$\{1+j,1-j,-1+j,-1-j\}$} and scaled by a small step size $\delta$, this divergence term can be \revb{estimated with $\mathcal{O}(\delta)$ bias} via
\begin{equation}
    \sum_{\mathbf{x}_{n}\in\mathcal{P}_{\mathrm{v}}}2\,\Re\Big\{\tfrac{\partial\hat{h}_{t}(\mathbf{x}_{n})}{\partial z_{n}}\Big\}\approx\frac{1}{\delta}\sum_{\mathbf{x}_{n}\in\mathcal{P}_{\mathrm{v}}}\Re\Big\{\big(\hat{h}_{t}^{+}(\mathbf{x}_{n})-\hat{h}_{t}(\mathbf{x}_{n})\big)\,b_{n}^{*}\Big\},
\end{equation}
where $\hat{h}_{t}^{+}(\cdot)$ denotes the output field under the perturbed inputs $\{z_{n}+\delta b_{n}\}$ and the computation requires only one additional forward pass per training step.

Notably, since the proposed loss in~\eqref{eq:gtfree-loss} depends exclusively on $\big(\{y_n\},\mathcal{P},\sigma_{w}^{2}\big)$, it can be evaluated directly on the live pilot stream without any labels. \revb{A model pretrained on an inaccurate offline simulator can therefore be continually refined via stochastic gradient descent on incoming pilot blocks, overcoming the distribution shift} identified in Section~\ref{sec:intro}.

\subsection{Optimality of the Proposed Loss}\label{sec:theory}

To formalize the theoretical guarantees \revb{of the proposed loss}, we call a candidate port \emph{scorable} if it falls into the masked set $\mathcal{P}_\mathrm{m}$ with positive probability conditional on every realization of the visible set $\mathcal{P}_\mathrm{v}$ that excludes it. \revb{While the existing self-supervised FAS estimator~\cite{gao2025ssnet} is justified only empirically, the following proposition shows that the proposed loss preserves the optimal MMSE solution despite relying solely on noisy sub-sampled pilots.}

\begin{proposition}\label{prop:mmse}
Assume that all candidate ports are scorable and that the FNO is sufficiently expressive to represent the ideal MMSE estimator
\begin{equation}
    \hat{h}_t^\star(\cdot) \triangleq \mathbb{E}\big[h(\cdot) \,\big|\, \{z_n\}, \mathcal{P}_\mathrm{v}, \sigma_t^2\big].
\end{equation}
Then, $\boldsymbol{\theta}$ minimizes $\mathcal{L}_\mathrm{prop}(\boldsymbol{\theta})$ if and only if the resulting output $\hat{h}_t$ coincides with $\hat{h}_t^\star$ at every candidate port.
\end{proposition}

Due to space limits, we only outline the core steps of the proof here. We first show that $\mathcal{L}_{\mathrm{prop}}(\boldsymbol{\theta})$ is an unbiased surrogate for the clean supervised loss $\mathcal{L}_{\mathrm{clean}}(\boldsymbol{\theta})\triangleq\mathbb{E}\big[\sum_{\mathbf{x}_{n}\in\mathcal{P}}\vert{}\hat{h}_{t}(\mathbf{x}_{n})-h(\mathbf{x}_{n})\vert{}^{2}\big]$. Expanding the squared error for each scored port yields
\begin{equation}
\begin{aligned}
    \vert{}\hat{h}_{t}(\mathbf{x}_{n})-y_{n}\vert{}^{2}&=\vert{}\hat{h}_{t}(\mathbf{x}_{n})-h(\mathbf{x}_{n})\vert{}^{2}\\
    &-2\Re\big\{(\hat{h}_{t}(\mathbf{x}_{n})-h(\mathbf{x}_{n}))w_{n}^{*}\big\}+\vert{}w_{n}\vert{}^{2}.
\end{aligned}
\end{equation}
Taking the expectation, the term $\vert{}w_{n}\vert{}^{2}$ evaluates to a constant $\sigma_{w}^{2}$. The behavior of the cross term depends on the visibility of the corresponding port. For masked ports $\mathbf{x}_{n}\in\mathcal{P}_{\mathrm{m}}$, $w_{n}$ is independent of both the network input and the channel, so the cross term vanishes. For visible ports $\mathbf{x}_{n}\in\mathcal{P}_{\mathrm{v}}$, $\hat{h}_{t}(\mathbf{x}_{n})$ correlates with $w_{n}$ via $z_{n}$. Applying Stein's lemma for complex Gaussians, the expected cross term becomes $-2\sigma_{w}^{2}\Re\big\{\mathbb{E}[\partial\hat{h}_{t}(\mathbf{x}_{n})/\partial z_{n}]\big\}$, which is exactly canceled by the divergence correction in~\eqref{eq:gtfree-loss}. Summing over the scored ports establishes $\mathcal{L}_{\mathrm{prop}}(\boldsymbol{\theta}) = \mathcal{L}_{\mathrm{clean}}(\boldsymbol{\theta}) + \mathbb{E}[M\sigma_{w}^{2}]$, where the second term is a constant independent of $\boldsymbol{\theta}$.

Next, expanding $\mathcal{L}_{\mathrm{clean}}(\boldsymbol{\theta})$ around the conditional mean $\hat{h}_{t}^{\star}(\cdot)$ yields $\mathcal{L}_{\mathrm{clean}}(\boldsymbol{\theta}) = \mathcal{L}_{\mathrm{MMSE}} + \mathbb{E}\big[\sum_{\mathbf{x}_{n}\in\mathcal{P}}\vert{}f(\mathbf{x}_{n})\vert{}^{2}\big]$, where $\mathcal{L}_{\mathrm{MMSE}}\triangleq\mathbb{E}\big[\sum_{\mathbf{x}_{n}\in\mathcal{P}}|\hat{h}_{t}^{\star}(\mathbf{x}_{n})-h(\mathbf{x}_{n})|^{2}\big]$ is the irreducible error of the MMSE estimator itself and $f(\cdot)\triangleq\hat{h}_{t}(\cdot)-\hat{h}_{t}^{\star}(\cdot)$ denotes the error field. Note that the cross term vanishes because the sampling pattern and split are both drawn independently of the channel realization. Therefore, minimizing $\mathcal{L}_{\mathrm{prop}}(\boldsymbol{\theta})$ reduces to minimizing $\mathbb{E}\big[\sum_{\mathbf{x}_{n}\in\mathcal{P}}\vert{}f(\mathbf{x}_{n})\vert{}^{2}\big]$. Finally, since the masked set is drawn independently of the channel and noise, the probability that a port hidden from the input is scored depends only on $\mathcal{P}_\mathrm{v}$ and is strictly positive by scorability. Hence, the above term vanishes if and only if $f$ vanishes at every candidate port, which is attainable under the expressiveness assumption and completes the proof.

\begin{remark}
Proposition~\ref{prop:mmse} guarantees optimality at all candidate ports of the grids on which the loss is evaluated. Its extension to the continuous aperture $\mathcal{D}$ as well as to unseen port resolutions relies on the resolution invariance property of the FNO, which stems from its grid-independent parameterization as seen in Section~\ref{sec:fno}. To reinforce this invariance, we adopt a multi-resolution training strategy that exposes the model to varying grid resolutions across coherence blocks, and validate this strategy empirically in Section~\ref{sec:experiments}.
\end{remark}

\section{Numerical Results}\label{sec:experiments}

In this section, we evaluate the proposed framework on channel realizations generated via the discrete multipath special case of~\eqref{eq:cont-channel} with $N_{p}=20$ paths, i.i.d. complex gains $g_{i}\sim\mathcal{CN}(0,1/N_{p})$, and AoAs uniformly distributed within maximum half angular spreads $\Theta=\Phi=60^{\circ}$. We consider a carrier frequency of $3.5$~GHz and an FAS aperture of size $L_{1}=L_{2}=4\lambda$, discretized into a square grid of $N_{\mathrm{r}} \times N_{\mathrm{r}}$ ports. The FNO architecture is configured with a mode truncation budget of $K_{1}=K_{2}=8$ and a feature dimension of $d=64$. With unit average channel power, the SNR is defined as $1/\sigma_w^2$. During training, the proposed estimator learns purely from noisy sub-sampled pilot streams, where the sampling patterns, SNRs, and pilot ratios are randomized per block. For pilot splitting, each training pilot block consists of evenly spaced ports, which serve as the visible set $\mathcal{P}_\mathrm{v}$, and $N/16$ additional ports drawn uniformly at random from the remaining ones, which form the masked set $\mathcal{P}_\mathrm{m}$. Moreover, to realize the multi-resolution strategy discussed in Section~\ref{sec:theory}, the training resolutions alternate among $N_{\mathrm{r}}\in\{24,36,48\}$.

We compare the proposed approach against the following baselines: (i)~a simple \revb{\textbf{linear interpolation}} baseline that fills the unobserved ports \revb{from} the noisy observations across the grid; (ii)~\revb{\textbf{OMP}} and (iii)~\revb{\textbf{SBL}}, which share a steering dictionary with resolution $G=50$ for both virtual azimuth and elevation AoAs; (iv)~\revb{\textbf{Supervised FNO}}, \revb{the} same FNO backbone trained on clean channel labels, which serves as an ideal performance reference; and (v)~\revb{\textbf{DDRM}, the diffusion-based estimator from our previous work~\cite{tang2025accurate} implemented via the denoising diffusion restoration model~\cite{kawar2022denoising}} with an unconditional prior trained on clean $48\times48$ channel samples. For this diffusion baseline, skipped sampling \revb{with $T=50$ out of $T=500$ full steps} is adopted for accelerated inference. All methods are evaluated on a common test set of $2048$ corrupted channel realizations.

\begin{figure}[t]
    \centering
    \includegraphics[width=0.85\columnwidth]{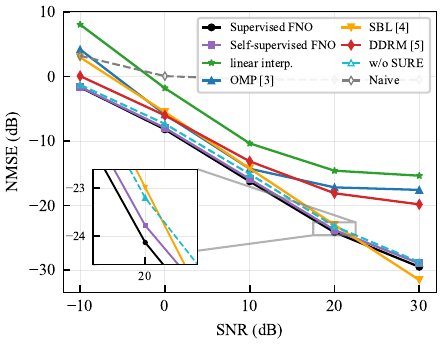}
    \caption{NMSE comparison under varying SNR.}
    \label{fig:snr-abl}
\end{figure}

\subsection{Estimation Accuracy and Ablation}\label{sec:exp-acc}

Fig.~\ref{fig:snr-abl} presents the normalized mean square error (NMSE) versus SNR at a resolution of $N_{\mathrm{r}}=48$ and a pilot ratio of $1/9$, where dashed curves denote the ablated variants of the proposed loss function. Notably, the FNO trained via the proposed self-supervised loss closely matches the supervised FNO across the entire SNR regime, empirically validating its optimality. Conversely, the naive loss~\eqref{eq:naive-loss} remains near $0$~dB regardless of the SNR, as it provides no supervision on the unobserved ports. Furthermore, omitting the SURE correction in~\eqref{eq:gtfree-loss} consistently degrades performance, particularly at low-to-moderate SNRs, due to overfitting to the measurement noise. These results confirm that both pilot splitting and SURE correction are essential components of the proposed loss.

Compared to the baselines, the proposed framework maintains a significant performance gain in the low-to-medium SNR regimes. Benefiting from a sparse parametric model matched to the simulated multipath channel, SBL outperforms the proposed method at $30$~dB. \revb{However, this advantage is tied to the exact sparsity of the simulated channel.} Finally, DDRM exhibits an error floor at high SNRs because under a tight pilot budget, its shortened reverse trajectory cannot provide sufficient measurement guidance for accurate recovery.

\begin{figure}[t]
    \centering
    \includegraphics[width=0.85\columnwidth]{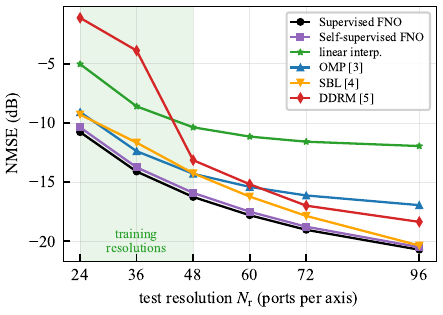}
    \caption{NMSE comparison under varying test resolution.}
    \label{fig:resolution}
\end{figure}

\subsection{Resolution Invariance}\label{sec:exp-res}

Fig.~\ref{fig:resolution} investigates the generalization of a single trained model across test resolutions $N_{\mathrm{r}} \in \{24, \dots, 96\}$ at $10$~dB SNR and a pilot ratio of $1/9$, where the resolutions beyond the maximum training grid of $N_{\mathrm{r}}=48$ are entirely unseen. As illustrated, the proposed self-supervised FNO closely matches its supervised counterpart across all tested resolutions, which confirms that the proposed self-supervised training scheme successfully preserves the inherent resolution invariance of the FNO architecture. Furthermore, while both FNO variants consistently outperform all other baselines, DDRM is structurally constrained to its fixed $48 \times 48$ training grid, leading to a severe performance collapse at coarser resolutions. \revb{Although OMP and SBL also support resolution invariance by operating in the virtual AoA domain, which is independent of the port grid, a significant accuracy gap from the FNO estimators persists across all tested resolutions.}

\begin{table}[t]
    \centering
    \caption{Online adaptation performance (NMSE in dB) \revb{versus the number of self-supervised adaptation updates}.}
    \label{tab:adaptation}
    \setlength{\tabcolsep}{3pt}
    \begin{tabular}{l>{\columncolor{gray!15}}cccccc}
        \hline
        \revb{Adaptation updates} & $0$ & $960$ & $1{,}920$ & $3{,}840$ & $7{,}200$ & $28{,}800$ \\
        \hline
        Training from scratch & -- & $-10.2$ & $-12.9$ & $-15.0$ & $-15.4$ & $-15.9$ \\
        \textbf{Proposed adaptation} & $\mathbf{-3.7}$ & $\mathbf{-11.1}$ & $\mathbf{-14.5}$ & $\mathbf{-15.4}$ & $\mathbf{-15.6}$ & -- \\
        \hline
    \end{tabular}
\end{table}

\subsection{Online Adaptation under \revb{Distribution Shift}}\label{sec:exp-adapt}

Finally, we validate the online adaptation capability introduced in Section~\ref{sec:loss}. We first pretrain a supervised model on a mismatched offline simulator \revb{with $\Theta=\Phi=30^{\circ}$ and $N_{p}=4$,} and then deploy it into the true environment \revb{with $\Theta=\Phi=60^{\circ}$ and $N_{p}=20$}. Table~\ref{tab:adaptation} tracks the test NMSE against \revb{the number of self-supervised updates performed on the live pilot stream} under $10$~dB SNR, $N_{\mathrm{r}} = 48$, and $1/9$ pilot ratio. While \revb{the highlighted zero-update column of Table~\ref{tab:adaptation} shows that deploying the pretrained model without any adaptation suffers severe degradation under the distribution shift}, the fully self-supervised nature of the proposed loss allows the estimator to continuously adapt on the incoming pilot stream. As shown in Table~\ref{tab:adaptation}, \revb{the proposed estimator successfully adapts online and converges much faster than training from scratch, almost halving the number of updates required to reach a similar NMSE}. These results confirm that \revb{the proposed self-supervised framework enables the estimator to recover from distribution shift using received pilots alone}.

\section{Conclusion}\label{sec:conclusion}

In this paper, we \revb{bridged the deployment gap between operator learning and practical channel acquisition with} a self-supervised training framework for \revb{FAS channel estimation}, which learns the \revb{FNO-based} estimator \revb{solely} from the noisy sub-sampled pilots available at the receiver. Specifically, \revb{the proposed} method integrates pilot splitting to supervise the network at hidden ports and applies a SURE correction to eliminate the measurement noise bias at visible ports. We \revb{proved} that under a sampling coverage condition, \revb{this proposed loss is minimized by the MMSE estimator, which coincides with the minimizer of the ideal supervised objective}. Simulation results demonstrated that the self-supervised FNO performs close to its supervised counterpart, preserves zero-shot generalization across port resolutions, and \revb{adapts online to distribution shift with almost half the updates required by training from scratch}. Future work will exploit the band-limited spatial spectrum of the channel field to extend the optimality guarantee beyond the training grids, and will evaluate the framework under diffuse scattering and on more realistic channel data.

\bibliographystyle{IEEEtran}
\bibliography{main}

\end{document}